\documentclass[aps,twocolumn,amsmath]{revtex4}

\renewcommand{\vec}{\boldsymbol}
\usepackage{graphicx}

\begin{document}
\title{Dissociation dynamics of a Bose-Einstein condensate of molecules} 
\author{Michael W. Jack and Han Pu}
\surname{Jack}
\affiliation{Department of Physics and Astronomy and Rice Quantum
Institute, Rice University, Houston, Texas 77251}

\date{\today}

\begin{abstract}
An unstable condensate of diatomic molecules will coherently disassociate into correlated 
pairs of atoms.  This dissociation process exhibits very rich quantum dynamics depending 
on the quantum statistics of the constituent atoms. We show that in the case of bosonic 
atoms Bose-enhancement can lead to stimulated dissociation, whereas, in the case of fermions 
Pauli-blocking of the available states and a build-up of coherence between molecules and atom 
pairs can give rise to incomplete dissociation of the molecules and transient 
association-dissociation oscillations.
\end{abstract}
\maketitle 
The ability to create quantum degenerate molecules composed of fermionic \cite{regal03,cubizolles03,strecker03} 
or bosonic \cite{herbig03,durr04} atoms by tuning a molecular level into resonance with the atomic 
states via a Feshbach resonance \cite{tiesinga93,inouye98,timmermans99} or by photoassociation \cite{fedichev96,fatemi00}  
has opened up an exciting new area of physics for exploration. The case of diatomic bosonic 
molecules coupled to bosonic atoms ($b \leftrightarrow bb$) has been shown to undergo coherent 
association-dissociation oscillations \cite{donley02} and there are predictions of Bose-enhanced  
phenomena in this system that may lead to a, so called, superchemistry \cite{heinzen00,kheruntsyan02,moore02,vardi02}.   
The case of diatomic bosonic molecules coupled to fermionic atoms ($b \leftrightarrow f\!f$) has 
also received a lot of attention lately due to the possibility of  realizing a BEC-BCS 
crossover \cite{leggett80,regal04,zwierlein04}. For positive detuning from resonance 
(where two-body theory predicts unstable molecules) molecules can be stabilized by Pauli-blocking 
of the atomic states \cite{falco04} and coherent population oscillations have also been predicted 
to occur in this case \cite{andreev04,barankov04,search04}. In these systems quantum statistics obviously 
play an important role and it is becoming clear that  atom-molecule coherence  generated by their 
coupling is one of the key elements to understanding their equilibrium and non-equilibrium 
behavior \cite{holland01,mackie02,andreev04,search04}.

In this paper we consider the production of correlated pairs of atoms by the spontaneous dissociation 
of a pure Bose-Einstein
condensate (BEC) of molecules \cite{greiner03,jochim03,zwierlein03,bourdel04}. This highly non-equilibrium, spontaneous-dissociation 
regime can be reached by  first creating a stable BEC of molecules far from the resonance, then 
rapidly tuning through to the other side of the resonance, i.e., $\nu<0 \rightarrow \nu>0$, where 
$\nu$ is the detuning of the molecular level from the atomic continuum and depends on the applied 
magnetic field in the case of a Feshbach resonance or the detuning of the coupling laser field in 
the case of photoassociation/dissociation. Once the molecule level is above the atomic continuum 
the molecules will become unstable and begin to dissociate into atomic pairs. For a condensate in 
a zero momentum state the atoms are created in correlated pairs of equal and opposite momentum 
centered at $k=\sqrt{2m\nu/\hbar}$.   Correlated pair production by this method has been discussed 
previously  in the case of bosonic atoms \cite{heinzen00,moore02}. Here we present a unified 
treatment of the dissociation dynamics of a molecular condensate which includes both the boson 
($b \rightarrow bb$) and fermion ($b \rightarrow f\!f$) case. 

Assuming the momentum spread of the molecular condensate is negligible compared to the mean momentum 
of the emitted atoms, the Hamiltonian of the system can be approximated by 
\begin{eqnarray}
H&=&\hbar\nu a^{\dagger}_{0}a_{0}+\hbar\sum_{\vec{k}}\omega_{\vec{k}}(b_{\vec{k}\uparrow}^{\dagger}
b_{\vec{k}\uparrow}+b_{\vec{k}\downarrow}^{\dagger}b_{\vec{k}\downarrow})\nonumber\\
&&+\hbar g\sum_{\vec{k}}\left(a_{0}^{\dagger}b_{-\vec{k}\downarrow}b_{\vec{k}\uparrow}+a_{0}
b^{\dagger}_{\vec{k}\uparrow}b^{\dagger}_{-\vec{k}\downarrow}\right)\,,\label{hamiltonian}
\end{eqnarray}
where $a_{0}$ is the bosonic molecular mode and $b_{\vec{k}\uparrow}$ and  $b_{\vec{k}\downarrow}$ 
are the
annihilation operators of the atoms and either satisfy bosonic or fermionic commutation relations. 
$\omega_{\vec{k}}=\hbar k^2/2m$ is the dispersion relation of the atoms and  $g$ is the coupling 
between the closed channel (molecules) and the open channel (free atoms) of the coupled channels 
scattering problem \cite{tiesinga93}. We have assumed that the atoms are in different internal 
states denoted by  $\uparrow$ and $\downarrow$ but the conclusions can be straight-forwardly 
applied to the case of only one bosonic atomic species. The total number of atoms, $N=2a^{\dagger}_{0} 
a_{0}+\sum_{\vec{k}} (n_{\vec{k}\uparrow}+n_{\vec{k}\downarrow})$  is conserved, where 
$n_{\vec{k}s}=b^{\dagger}_{\vec{k}s}b_{\vec{k}s}$ is the number operator for the atoms. As atoms 
of opposite spin and momentum are created and destroyed as pairs, the number difference, 
$n_{\vec{k}\uparrow}-n_{-\vec{k}\downarrow}$, is also conserved. 
In the boson case the correlation between the atoms created via molecular dissociation is analogous 
to that between photons created in a non-degenerate parametric amplifier (see  \cite{dansbook} and 
references within).  Quite recently, the pair correlations between
fermionic atoms with opposite momenta have been observed in the
noise spectrum of photodissociated cold molecules
\cite{greiner05}, following the theoretical proposal of
Ref.~\cite{altman04}. This pair correlation does not require the presence of a molecular condensate 
as it is simply a consequence of the form of the Hamiltonian and arises even in an incoherent 
dissociation process (such as from thermal molecules). On the other hand, only coherent molecular 
dissociation, such as from a condensate, can give rise to atom-molecule
 coherence and the related phenomena of coherent association-dissociation oscilations \cite{donley02}.  This coherence---characterized by a non-zero value of $\langle a_{0}^{\dagger}b_{-\vec{k}\downarrow}
b_{\vec{k}\uparrow}\rangle$---plays an important role in the present work. 

Given that the atoms are created in pairs, we can take advantage of a formal mapping between pairs 
of fermion operators and spin-$1/2$ Pauli matrices to write the Hamiltonian in a more natural form. 
This mapping has been exploited to determine the phase diagram of the BCS-BEC crossover and to 
predict non-equilibrium atom-molecule oscillations \cite{andreev04,barankov04,search04}. A similar mapping 
can be made for boson pairs and we treat the two cases in parallel.
We define new operators by: $\sigma_{\vec{k}-} =  b_{-\vec{k}\downarrow}b_{\vec{k}\uparrow}$, 
$\sigma_{\vec{k}+}= b^{\dagger}_{\vec{k}\uparrow}b^{\dagger}_{-\vec{k}\downarrow}$ and 
$\sigma_{\vec{k}z}  =  \frac{1}{2}(n_{\vec{k}\uparrow}+n_{-\vec{k}\downarrow}\mp 1)$.
It is easy to check that these operators satisfy the commutation relations: 
 $[\sigma_{\vec{k}\pm},\sigma_{\vec{k}^{\prime}\pm}]=0$, $[\sigma_{\vec{k}z} ,
\sigma_{\vec{k}^{\prime}+}] = \delta_{\vec{k},\vec{k}'}\sigma_{\vec{k}+}$, $[\sigma_{\vec{k}z} ,
\sigma_{\vec{k}^{\prime}-}] = -\delta_{\vec{k},\vec{k}'}\sigma_{\vec{k}-}$ 
 and 
\begin{equation}
[\sigma_{\vec{k}+} ,\sigma_{\vec{k}^{\prime}-}]=   \pm 2\delta_{\vec{k},\vec{k}'} \sigma_{\vec{k}z},\label{commutation}\end{equation}
where the upper (lower) sign  in Eq.~(\ref{commutation}) corresponds to fermions (bosons). The seemingly insignificant sign 
difference in the commutation relations in the two cases  leads to  completely different dynamics.

Writing the Hamiltonian (\ref{hamiltonian}) in terms of these new operators we have 
(minus a constant): \begin{equation}
H=\hbar\sum_{\vec{k}}\left[2(\omega_{\vec{k}}-\nu)\sigma_{\vec{k}z}+ g(a^{\dagger}_{0}
\sigma_{\vec{k}-}+\sigma_{\vec{k}+}a_{0})\right]\,, \end{equation}
and $N=2a^{\dagger}_{0}a_{0}+\sum_{\vec{k}}(2\sigma_{\vec{k}z}\pm 1)$.
 In the fermion case, this Hamiltonian describes an ensemble of independent  
two-level systems interacting with a single bosonic mode. The case of identical two-level systems: 
$\omega_{\vec{k}}=\omega_{0}$, is called the Dicke model and is an exactly solvable model that 
has been extensively studied in the quantum optics literature (see  Ref.\,\cite{chumakov96} and 
references within).

In the special case where the population in each mode is small, 
($\langle n_{\vec{k}\uparrow}+n_{\vec{k}\downarrow}\rangle \ll 1$), throughout the dissociation 
process, due to a large number of available states, we can make the approximation that  
$[\sigma_{\vec{k}-} ,\sigma_{\vec{k}^{\prime}+}]\approx  \delta_{\vec{k},\vec{k}'}$, 
independent of whether the operators describe bosons or fermions. In other words, the underlying 
quantum statistics of the atoms are unimportant and the Hamiltonian describes the coupling of a 
single mode to a continuum of bosonic modes with a quadratic dispersion relation. This model has 
been studied previously in the context of an atom laser \cite{hope97}. Under the Born-Markov 
approximation, which holds for weak coupling and large $\nu$ \cite{markovapprox}, the molecules 
will experience a rather trivial exponential decay and the final frequency distribution of the atom 
pairs will be the standard Lorentzian.

\begin{figure}
\begin{center}
\includegraphics[scale=0.4]{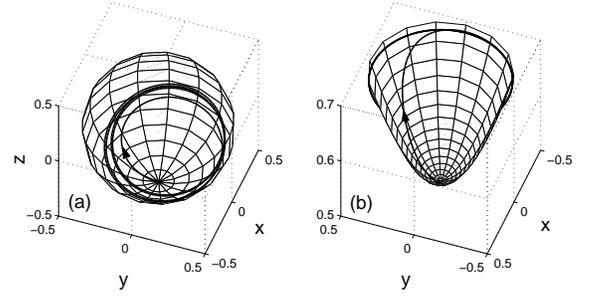}
\caption{\protect\label{trajectories} Example trajectories of $\vec{S}_{\vec{k}}(t)$ for fermions (a) and bosons (b).}
\end{center}
\end{figure}

To proceed further in the general case we note that we are interested in the spontaneous dissociation 
of an initially large molecular condensate, so we make a mean-field approximation for the molecular 
mode by replacing the operator $a_{0}(t)$ with a $c$-number $\alpha(t)$ (taken to be real without 
loss of generality).  This approximation will break down when the population of molecules approaches zero and quantum 
fluctuations of the molecular mode become important.  Under the mean-field approximation, the atom-molecule coherence reduces to 
$\langle a_{0}^{\dagger}b_{-\vec{k}\downarrow}b_{\vec{k}\uparrow}\rangle=\alpha\langle b_{-\vec{k}
\downarrow}b_{\vec{k}\uparrow}\rangle=\alpha\langle\sigma_{\vec{\vec{k}}-} \rangle$ and the  equations of motion 
for the averages $\langle \sigma_{\vec{k}s}\rangle$ can be written as:
\begin{equation}
\frac{d \vec{S}_{\vec{k}}}{dt} =\left(\begin{array}{ccc}
0& -2(\omega_{\vec{k}}-\nu)&0\\
2(\omega_{\vec{k}}-\nu)&0&\mp 2 g\alpha(t)\\
0& 2 g \alpha(t) &0
\end{array}
\right)\vec{S}_{\vec{k}}\,,\label{bloch}
\end{equation}
where $\vec{S}_{\vec{k}}=[\langle \sigma_{\vec{k}x}\rangle,\langle \sigma_{\vec{k}y}\rangle,\langle 
\sigma_{\vec{k}z}\rangle]^{t}$ is the column vector of averages  and 
we have defined $\sigma_{\vec{k}x}=(\sigma_{\vec{k}+}+\sigma_{\vec{\vec{k}}-})/2$ and 
$\sigma_{\vec{k}y}=(\sigma_{\vec{k}+}-\sigma_{\vec{k}-})/2i$.  In addition, $\alpha(t)$ is coupled 
to these variables via number conservation and is given by
\begin{equation}
\alpha(t)^{2}=N/2-\sum_{\vec{k}}[\langle\sigma_{\vec{k}z}(t)\rangle\pm \textstyle{\frac{1}{2}}]\,.
\label{alpha}
\end{equation}
Assuming an initial vacuum state for the atoms, $|{\rm vac}\rangle$, we have $\langle 
\sigma_{\vec{k}x}(0)\rangle=\langle \sigma_{\vec{k}y}(0)\rangle=0$ and $\langle \sigma_{\vec{k}z}(0)
\rangle=\mp \frac{1}{2}$. For fermions (upper sign) Eqs.~(\ref{bloch}) are the Bloch equations 
describing the dynamics of a two-level system driven by the classical field $\alpha(t)$. 
For each $\vec{k}$ the motion is confined to the surface of the  Bloch sphere defined by $\langle 
\sigma_{\vec{k}x}\rangle^2+\langle \sigma_{\vec{k}y}\rangle^2+\langle \sigma_{\vec{k}z}\rangle^2=1/4$ 
and is an expression of the underlying Fermi statistics (see Fig.~\ref{trajectories}). On the other hand, for bosons (lower sign) 
the motion is confined to the surface of a one-sided three dimensional (3D) hyperbola defined by 
$\langle \sigma_{\vec{k}z}\rangle^2-\langle \sigma_{\vec{k}x}\rangle^2-\langle \sigma_{\vec{k}y}
\rangle^2=1/4$ and  $\langle \sigma_{\vec{k}z}\rangle\geq 1/2$, and the population for each 
$\vec{k}$ is unbounded (see Fig.~\ref{trajectories}).

It is instructive to consider the case when $\alpha(t)=\alpha_{0}$ is a constant. 
In this case, Eq.~(\ref{bloch})  can be easily solved for the above initial state to yield
\begin{equation}
\langle \sigma_{\vec{k}z}(t)\rangle\pm\frac{1}{2} =\pm\frac{(g\alpha_{0})^{2}}
{2 \Omega_{\vec{k}\pm}^{2}}[1-C_{\vec{k}\pm}(t)]\,,\label{solution}
\end{equation}
where $\Omega_{\vec{k}\pm}^{2}=(g\alpha_{0})^{2}\pm(\omega_{\vec{k}}-\nu)^{2}$ and 
$C_{\vec{k}+}(t)=\cos(2\Omega_{\vec{k}+} t)$ and $C_{\vec{k}-}(t)=\cosh(2\Omega_{\vec{k}-} t)$. 
Again, the upper (lower) signs correspond to fermions (bosons). In the boson case 
this solution is unstable for $\alpha_{0}\neq 0$ as it leads to exponential growth 
at the rate $2\Omega_{\vec{k}-}$ for all $\Omega_{\vec{k}-}^{2}>0$. However, 
in the fermion case, the atom population is oscillatory  about the Lorentzian-shaped mean 
value: $(g\alpha_{0})^{2}/2\Omega_{\vec{k}+}^{2}$ and, in fact, we can find 
a self-consistent solution for $\alpha_{0}$ by substituting Eq.~(\ref{solution}) 
back into Eq.~(\ref{alpha}) and assuming  the oscillations eventually dephase for 
different $\omega_{\vec{k}}$ (this procedure is compared to the numerical solution in 
Fig.~\ref{Gamma100}). 
This solution is only valid when there is very little molecular decay, 
but it does indicate that a consistent solution can be found where the molecule 
population  does not completely decay away, and the numerical results presented below confirm this.

For illustrative purposes, we now confine our analysis to the case where $\nu$ 
is large so that the density of states is approximately flat across the region into which the 
molecules
tend to decay. 
In this case, any deviation from exponential decay can be directly attributed to the 
quantum statistics of the atoms rather than any structure in the density of states 
(c.f.  \cite{hope97}). Taking the continuum limit of Eq.~(\ref{alpha}) and evaluating 
the density of states at $\nu$  we can write
\begin{equation}
\alpha(t)^{2}\approx N/2-\rho(\nu)\int_{-\infty}^{\infty} d\delta [\langle\sigma_{z}
(\delta,t)\rangle\pm \textstyle{\frac{1}{2}}]\,,\label{numerical}
\end{equation}
where $\rho(\nu)=V\sqrt{\nu/2}\left(m/\hbar\right)^{3/2}/\pi^{2}$ is the density of states 
at $\nu$ for a uniform 3D box of volume $V$. Since the equation of motion for the $\sigma$'s 
only depends on $\delta=\omega_{\vec{k}}-\nu$ we have parameterized them by $\delta$ instead 
of $\vec{k}$.  As discussed above, in the regime where there are ample states available to 
the atoms, the molecules decay exponentially and the atoms populate the frequencies $\delta$ 
with a Lorentzian distribution of width $\sim g^{2}\rho(\nu)$. A measure of the number of 
available states is therefore given by this width multiplied by the density of states: 
$\rho(\nu)$. Motivated by this, we introduce the dimensionless quantity $\Gamma=N/[g\rho(\nu)]^2$,
such that the exponential-decay regime is given by $\Gamma\ll 1$. It follows that when this 
condition does {\em not} hold, we expect the behavior of the dissociation process to be 
altered by the quantum statistics of the atoms.

\begin{figure}
\begin{center}
\includegraphics[scale=0.4]{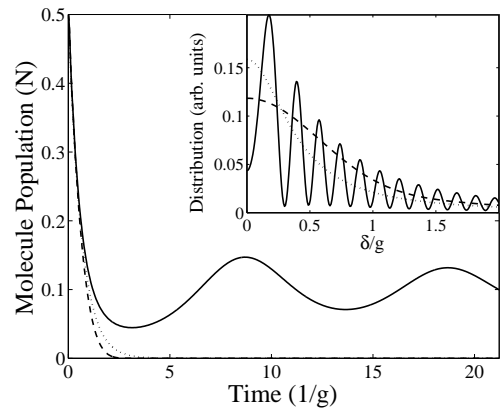}
\caption{\protect\label{Gamma4}Molecular population as a function of time for $\Gamma=4$  
in the case where the decay products are pairs of bosons (dashed line) and pairs of fermions 
(solid line). For comparison the usual exponential decay at the rate $\gamma=\pi\rho(\nu) g^{2}$ 
is shown (dotted line). Inset: normalized population distribution of atoms over 
$\delta=\omega_{\vec{k}}-\nu$ at the final time $tg=21.2$ for bosons (dashed line) and fermions 
(solid line). The dotted line is a Lorentzian of width $\gamma/4$. Only the $\delta\geq 0$ 
case is shown as the distribution is symmetric about $\delta=0$}
\end{center}
\end{figure}

\begin{figure}
\begin{center}
\includegraphics[scale=0.4]{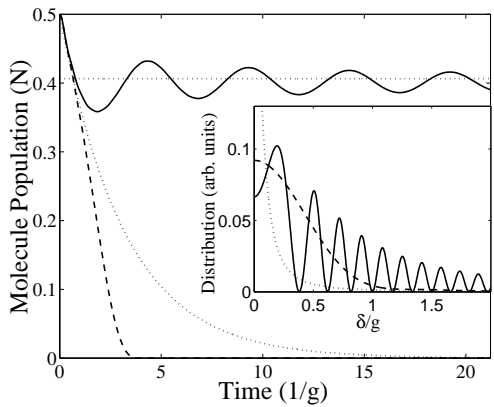}
\caption{\protect\label{Gamma100}Same as Fig.~\ref{Gamma4} but with $\Gamma=100$. The horizontal 
dotted line shows the steady-state molecular population determined by a self-consistent 
solution using Eqs.~(\ref{alpha}) and (\ref{solution})  (see text) which is a good fit in this case.}
\end{center}
\end{figure}

In Figure \ref{Gamma4} and \ref{Gamma100} we have plotted the results of a numerical solution 
of Eqs.~(\ref{bloch}) and (\ref{alpha}) for $\Gamma=4$ and $\Gamma=100$, respectively. Already 
for $\Gamma=4$ these plots show a marked deviation from the usual exponential decay. In particular, 
the molecular population in the fermion case does not decay to zero. For $\Gamma=100$ we see the 
accelerated decay that occurs due to bosonic stimulation in the boson case, whereas for the fermion 
case, a
large population remains in the molecular state.  

The behavior of the fermionic atoms can be qualitatively understood as follows: The molecular 
population initially undergoes a rapid decay into pairs of fermions. However, if $\Gamma\agt 1$, 
the states close to resonance, $\delta\approx 0$, become filled and begin to undergo coherent 
association-dissociation oscillations, effectively halting the molecular decay.  After a few 
oscillations the molecular population settles into a quasi-stationary state, $\alpha_{0}^{2}$ 
(which we have only been able to determine numerically), leaving the spin vectors $\vec{S}(\delta)$ 
precessing about the effective ``field'' $\vec{B}(\delta)=[2g\alpha_{0},0,2\delta]^{t}$, i.e. 
$\dot{\vec{S}}(\delta)=\vec{B}(\delta)\times \vec{S}(\delta)$. This results in the fringes in the 
distribution over $\delta$ (see insets in Figs.~\ref{Gamma4} and \ref{Gamma100}) which become more 
dense with time. The oscillation in the atomic population associated with this precession reacts 
back on the molecular field leading to an amplitude modulation of the  molecular population which 
damps slowly due to the dephasing of the spins with different $\delta$. In the final state the 
spins are completely dephased and no net dissociation or association can occur. This behavior is 
reminiscent of the processes of optical nutation, where an intense laser pulse excites an 
inhomogeneous media of two-level atoms \cite{allen75}. Unlike the soliton-like oscillations 
predicted in Refs.~\cite{andreev04}, the oscillations described here are a transient phenomena, 
but have the advantage that they can be created in a straight-forward manner, experimentally.
The incomplete dissociation for the fermion case was also found in Ref.~\cite{corney04} using a 
 stochastic wavefunction  approach.

Due to the assumption that the molecular condensate can be
described by a mean field the atom pairs with different $\vec{k}$ are uncoupled and
the Hilbert space of the system can be written as the tensor
product ${\mathcal H}=\prod_{\vec{k}}\otimes{\mathcal
H}_{\vec{k}}$, where ${\mathcal H}_{\vec{k}}$ is the Hilbert space
of the pair $(\vec{k}\!\!\uparrow,-\vec{k}\!\!\downarrow)$. We can
write the unitary evolution operator as
\begin{equation}
U_{\vec{k}}(t)=\exp\left\{r_{\vec{k}}(t)\left[e^{i\phi_{\vec{k}}(t)}
b_{\vec{k}\uparrow}^{\dagger}b_{-\vec{k}\downarrow}^{\dagger}-e^{-i\phi_{\vec{k}}(t)}
b_{-\vec{k}\downarrow}b_{\vec{k}\uparrow}\right]\right\}\,,\label{unitary}
\end{equation}
where $r_{\vec{k}}(t)$ and $\phi_{\vec{k}}(t)$ are real
time-dependent functions and are completely specified by the
expectation values $\langle \sigma_{\vec{k}s}(t)\rangle$.
Therefore, solving the Eqs.~(\ref{bloch}) enables us to determine
not only the expectation values $\langle
\sigma_{\vec{k}s}(t)\rangle$, but also the full quantum state of
the atoms. In the case of bosons, Eq.~(\ref{unitary}) is analogous
to the generator of a two-photon squeezed vacuum state
\cite{dansbook}. In the case of fermions, when
acting on the vacuum, this evolution operator yields the state
\begin{equation}
|\psi_{\vec{k}}(t)\rangle=\left[\cos r_{\vec{k}}(t)+
b_{\vec{k}\uparrow}^{\dagger}b_{-\vec{k}\downarrow}^{\dagger}e^{i\phi_{\vec{k}}(t)}
\sin r_{\vec{k}}(t)\right]|{\rm vac}\rangle,
\end{equation}
which  has obvious similarities to the BCS state. 

For short times we expect that the dynamics will be dominated by the $\vec{q}=0$ condensate of 
molecules. However, the $\vec{q}\neq 0$ molecular modes, collisions and finite temperatures that have been 
neglected from this treatment are expected to lead to a slow redistribution of the atoms and a 
decay of the correlations \cite{galitski05}. Determining the relaxation dynamics to the new 
equilibrium state is 
beyond the scope of the present work, however,
 we can understand some aspects of the relaxation dynamics by adding phenomenological phase damping 
terms to the equations of motion: $d \langle \sigma_{\vec{k}j}\rangle/dt|_{\rm relax}=- \langle 
\sigma_{\vec{k}j}\rangle/T_{2}'$, where $j=x,y$.
This relaxation process leads to a loss of the atom-molecule 
coherence ($\langle \sigma_{\vec{k}-}\rangle\rightarrow 0$) and will, in turn, destroy the dynamic 
equilibrium
reached in the absence of damping. This intuition is confirmed by
 numerical simulations which show that for finite
$T_{2}'$ the molecular population always decays to zero and clearly demonstrates the 
key role played by the atom-molecule coherence, in combination with Pauli blocking, 
in the incomplete
molecular dissociation in the fermion case.

 Due to their narrow dissociation linewidth,  narrow Feshbach resonances are promising 
systems to observe the  effects described here and  have the advantage that a magnetic 
field can be quickly tunned across the resonance into the dissociation regime.  In fact, 
the dissociation scheme considered here has recently been used to measure the width of a 
number of Feshbach resonances of  $^{87}$Rb \cite{durr04_sep}. For the $912$ G resonance 
with a width of $\Delta B=1.3$ mG, a detuning of $\hbar\nu=k_{B}\times 1\mu K$ yields a 
$\Gamma>3$ for typical densities ($n\sim 10^{13}$ cm$^{-3}$). Similarly, for the same 
parameters, the narrow $543$ G resonance of $^6$Li which has a width of 
$\Delta B=0.23$ G \cite{strecker03}, yields a $\Gamma>12$, demonstrating that the regime 
where quantum statistics play a role are well within the reach of current experiments.
Here we have assumed a uniform 3D
system but note that the effects of  quantum statistics can be significantly
enhanced in systems of reduced dimensionality\cite{moore02}, or in the presence
of trapping potentials, due to the reduction in the density of
states \cite{search02}.

 In summary, we have studied the effects of quantum statistics on
the dissociation dynamics of a condensate of diatomic molecules
formed either by two bosonic or fermionic atoms. In the former
case, the dissociation rate is Bose-enhanced;
while for the latter, Pauli-blocking in combination with the coherence formed between the  
molecules and atom pairs lead to a dynamic equilibrium between the molecule and
atom populations. Finally, we want to point out that we have used a method borrowed
from quantum optics, which can serve as a powerful tool to treat
other problems in the coupled atom-molecule system, such as the BEC-BCS
crossover.

We thank Randy Hulet for stimulating discussions which lead to this work.
HP acknowledges support from Rice University and Oak Ridge Associated Universities.


\end{document}